\documentclass{jetpl}
\usepackage{cite}
\twocolumn

\lat

\DeclareMathOperator{\sign}{sign}

\DeclareMathOperator{\Tr}{Tr}
\title{Josephson effect in S$_{\rm F}$XS$_{\rm F}$ junctions}

\rtitle{Josephson effect in S$_{\rm F}$XS$_{\rm F}$\ldots}

\sodtitle{Josephson effect in S$_{\rm F}$XS$_{\rm F}$ junctions}

\author{N.~M. Chtchelkatchev$^{+}$\/\thanks{e-mail:
nms@landau.ac.ru}, W. Belzig$^{*}$, and C. Bruder$^{*}$}

\rauthor{N.~M. Chtchelkatchev, W. Belzig, and C. Bruder}

\sodauthor{Chtchelkatchev, Belzig, and Bruder}

\address{$^+$L.D. Landau Institute for Theoretical Physics RAS,
117940 Moscow, Russia\\~\\
$^*$Departement Physik und Astronomie, Universit\"at Basel,
Klingelbergstrasse 82, 4056 Basel, Switzerland}

\dates{\today}

\abstract{We investigate the Josephson effect in S$_{\rm F}$XS$_{\rm
F}$ junctions, where S$_{\rm F}$ is a superconducting material with a
ferromagnetic exchange field, and X a weak link.  The critical current
$I_c$ increases with the (antiparallel) exchange fields if the
distribution of transmission eigenvalues of the X-layer has its
maximum weight at small values. This exchange field enhancement of the
supercurrent does not exist if X is a diffusive normal metal. At low
temperatures, there is a correspondence between the critical current
in an S$_{\rm F}$IS$_{\rm F}$ junction with collinear orientations of
the two exchange fields, and the AC supercurrent amplitude in an SIS
tunnel junction. The difference of the exchange
fields $h_1-h_2$ in an S$_{\rm F}$IS$_{\rm F}$ junction corresponds to
the potential difference $V_1-V_2$ in an SIS junction; i.e., the
singularity in $I_c$ [in an S$_{\rm F}$IS$_{\rm F}$ junction] at
$|h_1-h_2|=\Delta_1+\Delta_2$ is the analogue of the Riedel peak.
We also discuss the AC Josephson effect in S$_{\rm F}$IS$_{\rm F}$
junctions.
}


\begin{document}

\maketitle

The presence of a magnetic exchange field in bulk superconductors
\cite{Sarma,Buzdin} and in superconductor (S) -- ferromagnet(F)
multilayers reduces the critical temperature $T_c$, i.e., suppresses
superconductivity (see, e.g., \cite{Fominov} and references
therein). Similarly, an exchange field suppresses the proximity
effect: superconducting correlations spread into the F layer of
superconductor-ferromagnet structures on a shorter distance than into
the normal layer of a superconductor-normal metal structure
\cite{Demler}. Hence, it is natural to expect that the supercurrent in
a junction will be suppressed by an exchange field in the
superconductors or by the presence of ferromagnetic layers between the
superconducting banks. Surprisingly, it was shown recently that the
supercurrent can be strongly enhanced in a number of situations: e.g.,
in an S$_{\rm F}$IS$_{\rm F}$ junction formed by two ``ferromagnetic
superconductors'' (S$_{\rm F}$) whose exchange fields are oriented in
an antiparallel way \cite{Efetov,Krivoruchko}, and in SFIFS junctions
\cite{SFS_Bruder_Belzig,Kupriyanov}. There is still no simple
intuitive understanding of this exchange field supercurrent
enhancement (EFSE) effect, and also, which conditions favor this
effect. In the following, we investigate the Josephson effect in
S$_{\rm F}$XS$_{\rm F}$ junctions for different choices of the
scattering layer X, for example, when X is a diffusive normal metal or
an insulator, and find the conditions favoring the EFSE-effect.

In this Letter, we show that the EFSE-effect exists in S$_{\rm
F}$XS$_{\rm F}$ junctions if the distribution of transmission
eigenvalues of the X-layer has its maximum weight for small
values.
If the transparency is increased, we find that the effect becomes less
pronounced; it disappears when the transparency is close to unity. If
X is a diffusive normal metal, there is no exchange field enhancement of the
supercurrent. At zero temperature, we find a correspondence between
the critical current $I_c(V=0,h_1-h_2)$ of an S$_{F1}$IS$_{F2}$
junction with collinear exchange fields $h_{1(2)}$ and the AC
supercurrent amplitude $\Real I_c(V)$ of an SIS tunnel junction. The
two quantities coincide if the voltage $V$ across the junction is
equal to $h_1-h_2$. Thus, the peak-like singularity of
$I_c(V=0,h_1-h_2)$ at $|h_1-h_2|=\Delta_1+\Delta_2$ has the same
nature as the Riedel peak in SIS contacts at $|eV|=\Delta_1+\Delta_2$
\cite{Riedel,Werthamer,Larkin_Ovchinnikov,Kulik}. Here,
$\Delta_{1(2)}$ are the superconducting pair potentials of the two
contacts.
\begin{figure}
\begin{center}
\includegraphics[width=65mm]{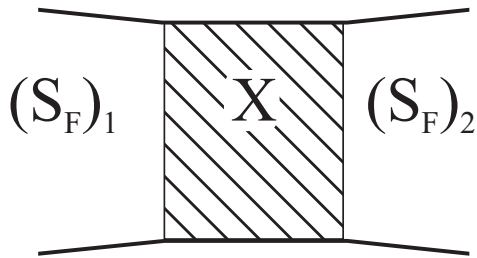}
\end{center}
\caption{\label{fig_1}\textbf{Fig. 1} Sketch of the device showing
the EFSE-effect; the two ferromagnetic superconductor layers
S$_{\rm F}$ are characterized by
BCS order parameters $\Delta_{1(2)}$ and exchange fields
$\mathbf{h}_{1(2)}$. A scattering region X (e.g., an insulator or
a diffusive normal metal) separates the two S$_{\rm F}$ layers.}
\end{figure}

To derive the results listed above, we relate the supercurrent
through the S$_{\rm F}$XS$_{\rm F}$ junction to the scattering
matrix of the region X, and then use the statistical properties of
this scattering matrix. The model considered is illustrated in
Fig.~\ref{fig_1}. It consists of a scattering region (hatched)
between two superconducting S$_{\rm F}$ layers.

Examples for S$_{\rm F}$-layers include superconductors with
ferromagnetic impurities \cite{Sarma}, or
superconductor-ferromagnet (normal metal) multilayers, where the
superconducting (and ferromagnetic) order parameter is induced by
the proximity effect \cite{Efetov,Nazarov}. They can be described
by adding an exchange field to the BCS-model
\cite{abrikosov,fominov_prb}. Then the self-consistency equation at
zero temperature shows that the superconducting order parameter
$\Delta(h)=\Delta(0)$ if the exchange field $h<\Delta(0)$, and
$\Delta(h)=0$ otherwise. In this paper we assume that $|h|\leq
\Delta(0)$ in the two ``ferromagnetic superconductor'' leads.

The supercurrent is calculated using the quasiclassical Green's
function technique. We assume that the junction is short, i.e.,
that the traversal time $\tau$ through the region X is such that
$\hbar/\tau$  exceeds the superconducting order parameters
$\Delta_{1,2}$ of the S$_{\rm F}$ layers. Then, following
\cite{Zaitsev,circuittheory}, we relate the supercurrent $I$ to
Keldysh Green's functions, and finally to retarded quasiclassical
Green's functions $\hat R_{1,2}$ in the bulk of the S$_{\rm F}$
layers, and the eigenvalues ${\mathcal{T}}_n$ of $tt^\dag$, where
$t$ is the transmission amplitude of the X-layer:
\begin{subequations}
\begin{eqnarray}
\label{I} I&=&\frac 1{4e} \int dE\,{\mathrm{Tr}} \left[\hat
\tau^{(3)}\hat I(E)\right] \tanh\left(\frac E {2T}\right)\; ,
\\
\label{I_general} \hat I &=& \frac {e^2} {\pi\hbar} \sum_n
2{\mathcal{T}}_n \frac{[\hat R_1,\hat
R_2]}{4+{\mathcal{T}}_n(\{\hat R_1,\hat R_2\}-2)}\; ,
\\
\label{R} \hat
R_{1,2}(E)&=&\frac{i}{\sqrt{(\Delta_{1,2})^2-(E+\vec\sigma
\cdot{\mathbf{h}}_{1,2})^2}}\times\\ \notag
&&\times\left(%
\begin{array}{cc}
 E+\vec\sigma \cdot {\mathbf{h}}_{1,2} &
 e^{i\varphi_{1,2}}\Delta_{1,2} \\
 -e^{-i\varphi_{1,2}}\Delta_{1,2} &
 -E-\vec\sigma \cdot {\mathbf{h}}_{1,2} \\
\end{array}%
\right)\; .
\end{eqnarray}
\end{subequations} Here, $\hat \tau^{(3)}$ is the Pauli matrix
acting in Nambu space, the trace is taken over the Nambu and
spin-spaces, and $\varphi_{1,2}$ is the superconducting phase
corresponding to the S$_{\rm F}$ layers.
Equations~(\ref{I}-\ref{R}) are valid for both ballistic or dirty
S$_{\rm F}$-layers.

\begin{figure}
\begin{center}
\includegraphics[width=75mm]{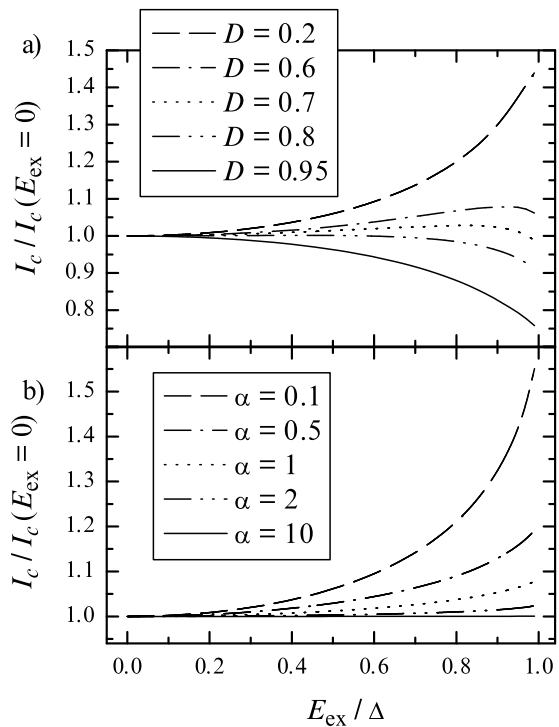}
\end{center}
\caption{\label{Fig:3}\textbf{Fig. 3} Exchange field dependence of
the critical current in an S$_{\rm F}$XS$_{\rm F}$ junction with
$\Delta_1=\Delta_2$ and $E_{\rm ex}\equiv h_1=-h_2$. (a) X is an
insulator with transparency $D$. For $D>0.7$, the supercurrent
enhancement effect disappears. (b) X is a disordered normal metal
of conductance $G_N$ with a tunnel junction of conductance $G_T$;
$\alpha=G_T/G_N$. The supercurrent enhancement effect disappears
for $\alpha\gg 1$.}
\end{figure}

To derive Eqs.~(\ref{I},\ref{I_general}) we used the general
Zaitsev boundary conditions \cite{Zaitsev,circuittheory} for the
Green's functions rather than the Kupriyanov-Lukichev dirty-limit
approximation \cite{Kuprianov-lukichev} which is valid for small
$\cal T$ (see, e.g., \cite{Belzig} and references therein).  Using
the Zaitsev boundary condition leads to the anticommutator of the
Green's functions in the denominator of Eq.~(\ref{I_general}) which
plays an important role here and cannot be neglected.  Due to this
anticommutator the EFSE-effect is suppressed in S$_{\rm
F}$IS$_{\rm F}$ junctions with large transparencies $\cal T$ and
in S$_{\rm F}$XS$_{\rm F}$ junctions in which X is a dirty normal
metal, see, e.g., Fig.~\ref{Fig:3}.

If ${\mathbf h}_1 \| \, {\mathbf h}_2$, Eq.~\eqref{I_general}
reduces to:
\begin{gather}
\label{I_ex} I(\varphi)= \sum_{\sigma=\pm 1}\int d{\cal
T}\rho({\cal T})\frac{e}{\hbar}T\sum_\omega \frac d
{d\varphi}\ln[g(i\omega,\varphi, \sigma,{\cal T})]\; .
\end{gather}
Here,
\begin{multline}
\label{g}
g(E,\varphi,\sigma,{\cal T})= (1-{\cal T})\sin(a_1)\sin(a_2)+ \\
\frac 1 2 {\cal T}(\cos\left(\varphi\right)-\cos(a_1+a_2))\; ,
\end{multline}
where $\varphi=\varphi_1-\varphi_2$, $\rho({\cal
T})=\sum_n\delta({\cal T}-{\cal T}_{n})$ is the distribution of
transmission eigenvalues, $\omega=2\pi T(k+1/2),\,k=0,\pm
1,\ldots$ are Matsubara frequencies, and
$a_{1,2}=\arccos[(E+\sigma h_{1,2})/\Delta_{1,2}]$ represent the
phases picked up at an Andreev reflection from the S$_{\rm F}$
layers.
Equations~\eqref{I_ex} and \eqref{g} can be also derived using the
scattering theory developed in Ref.~\cite{beenakker}.

In the general case ${\mathbf h}_1\nparallel{\mathbf h}_2$, the
supercurrent is given by
\begin{gather}
\label{I_theta} I(\varphi)=I^{(p)}(\varphi)\cos^2\left(\frac
{\theta} 2\right)+I^{(a)}(\varphi)\sin^2\left(\frac {\theta}
2\right)\; ,
\end{gather}
where the indices $p,a$ correspond to the parallel and
antiparallel configurations of the exchange fields; $\theta$ is
the angle between $\mathbf{h}_1$ and $\mathbf{h}_2$. Equation
\eqref{I_theta} can be derived from \eqref{I}-\eqref{R} using
the following identity for an analytical function $L$ of two
variables:
\begin{multline}
\label{lemma} \Tr L[(\vec \sigma\cdot{\bf a}),(\vec \sigma\cdot{\bf
  b})] \equiv\\ \frac 1 2\sum_{\sigma_{1(2)}=\pm
1}\left(1+\sigma_1\sigma_2\frac{{\bf a}\cdot{\bf b}}{|{\bf
a}||{\bf b}|}\right )L[\sigma_1|{\bf a}|,\sigma_2|{\bf b}|]\; ,
\end{multline}
where the trace is taken over spin degrees of freedom.
The last identity can be proved by a series expansion.

Using Eqs.~(\ref{I_ex}-\ref{I_theta}), we can
work out the effect of ferromagnetic interactions on the
supercurrent in a number of structures.

We shall concentrate below on the case when the exchange fields
are collinear. Suppose that X is a tunnel barrier. Then
$\rho({\cal T})=N\delta({\cal T}-D)$, where $D\ll 1$, $N$ is the
number of channels [$N=k_F^2A/4\pi$, where $A$ is the area of the
junction cross-section, and $k_F$ is the Fermi wave vector in S].
It follows from \eqref{I_ex} that
\begin{gather}
\label{I_T}
I(\varphi)=\sin(\varphi)\frac{\pi}{e}(R_N)^{-1}\Delta_1\Delta_2\times \\
\nonumber T\sum_{\omega}\Real \frac 1
{\sqrt{(\Delta_1)^2+(\omega+ih_{1})^2}\sqrt{(\Delta_2)^2+
(\omega+ih_{2})^2}}\; ,
\end{gather}
where $R_N=(NDe^2/\pi\hbar)^{-1}$ is the normal-state resistance of the
junction. If $\sign (h_1 h_2)>0$, Eq.~\eqref{I_T} gives
$I^{(p)}$, and in the opposite case  $I^{(a)}$ [see
Eq.~\eqref{I_theta}]. For $\Delta_1=\Delta_2$,
$h_1=-h_2$, Eq.~\eqref{I_T} reproduces the corresponding results of
\cite{Efetov}.

\begin{figure}
\begin{center}
\includegraphics[width=75mm]{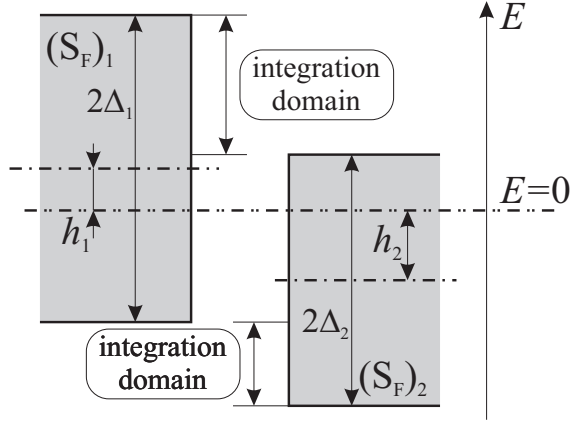}
\end{center}
\caption{\label{fig:2}\textbf{Fig. 2} The integration domain shown
gives the main contribution to the supercurrent in an (S$_{\rm
F}$)$_1$I(S$_{\rm F}$)$_2$ junction according to
Eq.~(\ref{I_R_real_E}). The supercurrent shows a
 Riedel singularity when $|h_1|\to\Delta_1$, $|h_2|\to\Delta_2$.}
\end{figure}

It follows from Eq.~\eqref{I_T} that at small temperatures, $T\ll
\min\{\Delta_1, \Delta_2\}$, as long as $|h_1|<\Delta_1,
|h_2|<\Delta_2$, the supercurrent does not depend on $h_1+h_2$.
It grows with $h_1-h_2$ and diverges logarithmically when
$|h_1-h_2|\to \Delta_1+\Delta_2$. To illustrate this, we
write Eq.~\eqref{I_T} in the real-time representation:
\begin{gather}
\label{I_R_real_E} I(\varphi)=\frac{\Delta_1\Delta_2}{4
R_N}\sin(\varphi)\sum_{\sigma=\pm 1}\int_{-\infty}^\infty
dE\tanh\left(\frac E {2T}\right)\times \\
\Imag\frac{1}{\sqrt{((\Delta_1)^2
-(E+h_1\sigma)^2)((\Delta_2)^2-(E+h_2\sigma)^2)}}\; .
\notag
\end{gather}
The integration domain is shown in Fig.~\ref{fig:2}.
Equation~\eqref{I_R_real_E} and Fig.~\ref{fig:2} show that the
exchange fields $h_{1(2)}$ shift the Fermi energies of the two
superconductors by $\sigma h_{1(2)}$. The potentials $V_{1(2)}$
applied to the superconducting banks of an SIS junction shift the
Fermi energies in a similar manner. In particular, it turns out
that the amplitude $\Real I_c(V)$ of the AC Josephson supercurrent
[which is proportional to $\sin(2eVt/\hbar)$] of an SIS junction
is equal to the critical current $I_c=I(\varphi=\pi/2)$ in
Eqs.~(\ref{I_T},\ref{I_R_real_E}) after the substitution
$h_{1(2)}\to eV_{1(2)}$.
At zero temperature, the critical current $I_c=I(\varphi=\pi/2)$
defined by Eq.~\eqref{I_R_real_E} can be expressed through the
elliptic function $\mathbf K$
\cite{Werthamer,Larkin_Ovchinnikov,Gradshtein}. If we define
$h\equiv h_1-h_2$, then within the interval
$|h|<|\Delta_1-\Delta_2|$,
\begin{multline}
 I_c R_N
  =\frac{2e\Delta_1\Delta_2}{\sqrt{(\Delta_1+\Delta_2)^2-h^2}}\times\\
  {\mathbf K}\left(\sqrt{\frac{(\Delta_1-\Delta_2)^2-h^2}
  {(\Delta_1+\Delta_2)^2-h^2}}\right)\; .
\end{multline}
If $|\Delta_1-\Delta_2|<|h|<\Delta_1+\Delta_2$ then
\begin{gather}
\label{I_c_h0} I_c R_N =e\sqrt{\Delta_1\Delta_2}\,{\mathbf K}
\left(\sqrt{\frac{4\Delta_1\Delta_2}{h^2-(\Delta_1-\Delta_2)^2}}\right)\;
.
\end{gather}
For $h_1=h_2=0$, $\Delta_1=\Delta_2$,
Eq.~\eqref{I_c_h0} leads to $I_cR_N=e\Delta\pi/2$, i.e., the usual result
of the critical current of an SIS Josephson junction \cite{Kulik}.

For $|h|$ close to $\Delta_1+\Delta_2$, the integral
\eqref{I_R_real_E} has a singularity. The singular part of the
current is:
\begin{gather}
I_c R_N\sim\frac{e\sqrt{\Delta_1\Delta_2}} 2
\ln\left(\frac{\Delta_1+\Delta_2}{||h|-(\Delta_1+\Delta_2)|}\right)\;
.
\end{gather}

If the temperature is close to the critical temperature of the
S$_{\rm F}$ layer, the supercurrent depends on $h_1+h_2$ as well
as $h_1-h_2$, and there is no EFSE-effect in agreement with
\cite{Efetov}. In this case, the correspondence of the exchange
field in S$_{\rm F}$XS$_{\rm F}$ junctions and the voltage in SIS
junctions is not valid any more.

The main point of the discussion above is that the supercurrent is
strongly enhanced by the exchange field in the tunnelling regime,
i.e., when the scattering region X is an insulator with small
transparency. Below we investigate whether
the enhancement effect is seen in other types of S$_{\rm
F}$XS$_{\rm F}$ junctions, e.g., when the layer X is a diffusive normal
metal.

If $\Delta=\Delta_1=\Delta_2$, $h\equiv h_1=-h_2$ (antiparallel
magnetizations), Eq.~\eqref{g} can be simplified:
\begin{gather}
g(E,\varphi,\sigma,{\cal T})
=\frac{2-{\cal T}}{2\Delta^2}\sqrt{(\Delta^2-E^2-h^2)^2-4E^2h^2}+\notag
\\\label{g_simple}
\frac{\cal
T}{2}\left(\cos\left(\varphi\right)+\frac{h^2-E^2}{\Delta^2}\right)\; .
\end{gather}
The current can be evaluated using \eqref{I_ex}.

Let us first turn to the case when the distribution of transmission
eigenvalues $\rho\propto\delta({\cal T}-D)$.  As shown above, the
enhancement effect exists as long as $D\ll 1$. If the transparency $D$
becomes larger, we find from Eq.~\eqref{I_ex} that the EFSE-effect
becomes less pronounced; it disappears when the transparency is close
to unity.
This is illustrated in Fig.~\ref{Fig:3}a, where the critical
current of an S$_{\rm F}$XS$_{\rm F}$ junction with $\Delta\equiv
|\Delta_1|=|\Delta_2|$ is shown as a function of the exchange
field $E_{\mathrm{ex}}\equiv h_1=-h_2$ at different transparencies
$D$. The relation of the transparency and the normal state
resistance is given by $D=R_{\rm Sh}/R_N$, where the Sharvin
resistance $R_{\rm Sh}=(e^2k_F^2 A/4\pi^2\hbar)^{-1}$, and $A$ is
the area of the junction.

Another possibility is that X is a dirty normal wire of
conductance $G_N$ and an insulating layer with conductance $G_T$
crosses the wire [this insulating layer, for example, can be
situated at the S$_{\rm F}$-X interface]. In this case the
distribution of the transmission eigenvalues $\rho({\cal T})$ is
known \cite{Nazarov_2}; for example, if $G_T/G_N \gg 1$, then
$\rho({\cal T})= (\pi\hbar G_N/e^2)/{\cal T}\sqrt{1-{\cal T}}$
\cite{Dorokhov}. The graph of the critical current versus the
exchange field is shown in the Fig.~\ref{Fig:3}b for a set of
values of $\alpha\equiv G_T/G_N$. It follows from this figure that
in the metallic regime, $\alpha\gg 1$, when both small and large
transmission eigenvalues give the main contribution to the
current, the EFSE is suppressed. If X consists of two insulating
barriers separated by a dirty normal wire, $\rho\propto1/{\mathcal
T}^{3/2}\sqrt{1-{\mathcal T}}$; there is a weak EFSE-effect in
this case, the relative supercurrent enhancement does not exceed
10$\%$.

\begin{figure}
\begin{center}
\includegraphics[width=75mm]{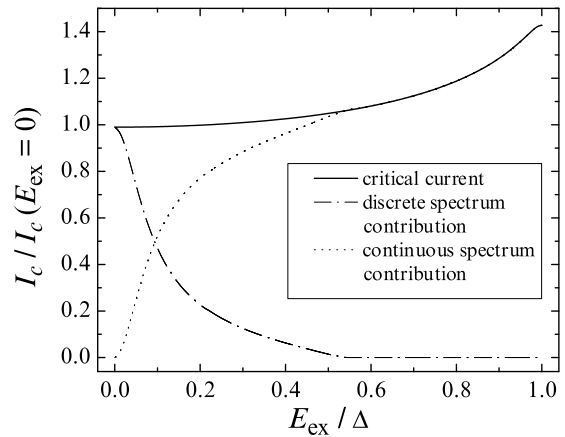}
\end{center}
\caption{\label{Fig:4}\textbf{Fig. 4}  The critical current in an
S$_{\rm F}$XS$_{\rm F}$ junction with $\Delta_1=\Delta_2$, $E_{\rm
ex}\equiv h_1=-h_2$, $\rho({\cal T})\propto \delta({\cal T}-D)$,
and $D=0.2$. The figure shows the relative contributions to the
critical current of the discrete spectrum (Andreev levels) and the
continuous spectrum.}
\end{figure}

Figure~\ref{Fig:4} shows the relative contribution of the discrete
spectrum (Andreev levels) and the continuous spectrum to the
supercurrent.  It turns out that the EFSE-effect is mostly due to
the continuous spectrum: the contribution of the discrete spectrum
to the supercurrent decreases with the exchange field, while the
contribution of the continuous spectrum increases. If X is an
insulator, the continuous spectrum gives the main contribution to
the supercurrent (see Fig.~\ref{fig:2}) and there is a pronounced
EFSE effect.

\begin{figure}
\begin{center}
\includegraphics[width=75mm]{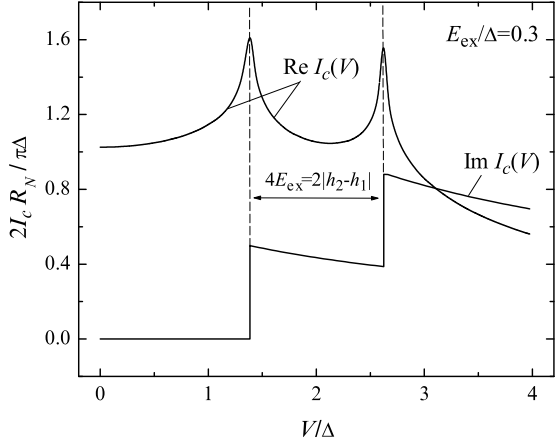}
\end{center}
\caption{\label{fig:5}\textbf{Fig. 5} Real and imaginary parts of
the AC Josephson supercurrent amplitude $I_c(V)$ in an S$_{\rm
F}$IS$_{\rm F}$ junction at $T=0$, $E_{\mathrm{ex}}\equiv
h_1=-h_2=0.3\Delta$, and $\Delta_1=\Delta_2\equiv\Delta$.
Riedel-type
singularities are visible at  $V=2\Delta\pm 2E_{\mathrm{ex}}$.}
\end{figure}

Finally we discuss the AC Josephson effect in S$_{\rm F}$IS$_{\rm
F}$ structures. Similarly as in tunnel SIS junctions
\cite{Werthamer,Larkin_Ovchinnikov,Kulik}, the current consists
of three parts: $I(t)=I_1(t)+I_2(t)+I_3$, where
$I_1(t)=\Real[I_c(V,h)]\sin(2eVt/\hbar)$ is the supercurrent,
$I_2(t)=\Imag[I_c(V,h)]\cos(2eVt/\hbar)$ the interference current,
and $I_3$ the quasiparticle current, here, $h=h_1-h_2$. Here, we
concentrate on the behavior of $I_1$ and $I_2$; the quasiparticle
current has been studied in \cite{Nazarov}. The complex
supercurrent amplitude $I_c(V,h)$ in an S$_{\rm F}$IS$_{\rm F}$
junction can be calculated in a similar way as in an SIS junction
\cite{Werthamer,Larkin_Ovchinnikov}. At zero temperature, it has
the remarkable property that
\begin{gather}
I_c(V,h)=\frac{1}{2}(I_c(V+h/e,0)+I_c(V-h/e,0))\; .
\label{icvh}
\end{gather}
By setting $V=0$, we find again that the DC critical current of an
S$_{\rm F}$IS$_{\rm F}$ junction coincides with the real part of
the AC supercurrent amplitude of an SIS junction if we replace
$eV$ by $h$. Using Eq.~(\ref{icvh}) we can also discuss the AC
Josephson effect of the S$_{\rm F}$IS$_{\rm F}$ junction. In an
SIS junction, $\Real I_c(V)$ has a Riedel singularity at
$|eV|=\Delta_1+\Delta_2$; but in the S$_{\rm F}$IS$_{\rm F}$ case,
the Riedel singularity appears at $|eV\pm
(h_2-h_1)|=\Delta_1+\Delta_2$ (we assume a collinear orientation
of the exchange fields $h_{1,2}$). In an SIS junction, $\Imag
I_c(V)$ vanishes for $|eV|<\Delta_1+\Delta_2$, and jumps to
$\pi\sqrt{\Delta_1\Delta_2}/2R_N$ at
$|eV|=\Delta_1+\Delta_2$\cite{Kulik}. In contrast, in an S$_{\rm
F}$IS$_{\rm F}$ junction, $\Imag I_c(V)$ jumps at
$|eV-(h_2-h_1)|=\Delta_1+\Delta_2$ [see Fig.~\ref{fig:5}], and the
the jump is half as big as in the SIS case.

In conclusion, we have shown that there is a pronounced exchange field
supercurrent enhancement effect in S$_{\rm F}$XS$_{\rm F}$ junctions
if the distribution of transmission eigenvalues of the X-layer has
maximum weight at small values. If X is a diffusive normal metal,
there is no exchange field enhancement of the supercurrent. At small
temperatures, there is a correspondence between the critical current
in an S$_{\rm F}$IS$_{\rm F}$ junction with collinear orientations of
the exchange fields and the supercurrent amplitude in an SIS tunnel
junction in the AC regime: the difference of the exchange fields in an
S$_{\rm F}$IS$_{\rm F}$ junction is the analogue of the voltage in an
SIS junction.  Finally, we also discussed the AC Josephson effect in
S$_{\rm F}$IS$_{\rm F}$ junctions.

We would like to thank V.~V. Ryazanov, Ya. Fominov, A.~A. Golubov,
A. Iosselevich, M.~V. Feigelman, and M. Skvortzov for stimulating
discussions and useful comments on the manuscript. The research of
N.M.C. was supported by the RFBR (projects No. 00-02-16617,
02-02-16622, and 02-02-06509), by Forschungszentrum J\"ulich
(Landau Scholarship), by the Netherlands Organization for
Scientific Research (NWO), by the Swiss NSF, and by the Russian
Ministry of Science (project \textit{Mesoscopic systems}). W.B.
and C.B. would like to thank the Lorentz Center in Leiden where
this manuscript was finished, and the Swiss NSF and the NCCR
Nanoscience for financial support.

\end{document}